# Realization of Room-Temperature Phonon-limited Carrier Transport in Monolayer MoS$_2$ by Dielectric and Carrier Screening

*Zhihao Yu, Zhun-Yong Ong, Yiming Pan, Yang Cui, Run Xin, Yi Shi\*, Baigeng Wang, Yun Wu, Tangsheng Che[4], Yong-Wei Zhang, Gang Zhang\* & Xinran Wang\**

Zhihao Yu, Yang Cui, Run Xin, Prof. Yi Shi, Prof. Xinran Wang
National Laboratory of Solid State Microstructures, School of Electronic Science and Engineering, and Collaborative Innovation Center of Advanced Microstructures
Nanjing University, Nanjing 210093, P. R. China
E-mail: xrwang@nju.edu.cn, yshi@nju.edu.cn
Zhun-Yong Ong, Yong-Wei Zhang，Gang Zhang
Institute of High Performance Computing, 1 Fusionopolis Way, 138632, Singapore
E-mail: zhangg@ihpc.a-star.edu.sg
Yiming Pan, Prof. Baigeng Wang
National Laboratory of Solid State Microstructures, School of Physics, Nanjing University, Nanjing 210093, P. R. China
Yun Wu, Tangsheng Chen
Science and Technology on Monolithic Integrated Circuits and Modules Laboratory, Nanjing Electronic Device Institute, Nanjing, China

Research into the physical properties of MoS$_2$ and other semiconducting transition metal dichalcogenides[1] (TMDs) has increased considerably in recent years, owing to their potential applications in post-CMOS electronics[2-4], optoelectronics[5-7] and valleytronics[8-10]. Some of the properties of monolayer MoS$_2$ that are advantageous for electronic applications include a direct band gap of 1.8 eV[11,12] as well as a film thickness of less than 1 nm which gives superior electrostatic control of the charge density and current even at the transistor scaling limit[3,13]. In spite of these favorable properties, the widely reported low electron mobility in monolayer MoS$_2$ poses a serious obstacle to its integration into post-CMOS nanoelectronics. The nature of charge transport in MoS$_2$, especially at room temperature, remains poorly understood despite considerable amount of theoretical and experimental researches. For example, the theoretically predicted intrinsic phonon-limited mobility at room temperature is in the range of 200-410 cm$^2$/Vs[14,15] while most experimentally reported values are much smaller[16-23]. Before any semiconducting material can become useful for potential nanoelectronic device applications, a critical assessment of its intrinsic charge transport properties at room temperature is needed, requiring the realization of high-quality samples with carrier mobility in the phonon-limited regime. The phonon-limited transport regime was demonstrated for graphene[24] and carbon nanotubes[25]. However, despite many recent efforts to improve carrier mobility by means of topgate[17], chemical functionalization[21] and BN gate dielectrics[22], phonon-limited transport regime has not been explicitly demonstrated in monolayer TMDs including MoS$_2$.

The possible reasons for the discrepancy between the theoretical upper limit and experimental data include Coulomb impurities (CI), traps and defects in low-quality samples[19-23]. These extrinsic sources of scattering have so far precluded any rigorous examination of the intrinsic scattering mechanisms that affect electron mobility. A particularly important source of scattering is from CI at the semiconductor-dielectric interface, which is believed to be the most important limiting factor in current MoS$_2$ devices[26]. Recently, it has been demonstrated that by sandwiching the monolayer MoS$_2$ channel between BN layers, CI scattering can be significantly suppressed, leading to a record-high mobility of over 1000 cm$^2$/Vs at low temperatures[22-23]. The technologically relevant room-temperature mobility, however, still lags the best devices on SiO$_2$ for reasons not well understood. Nonetheless, significant recent progress in reducing the deleterious effects of CI, traps and defects on the mobility[20-23] begin to set the stage for the realization of room-temperature charge transport in the phonon-limited regime.

It has been shown experimentally that the deposition of a high-κ top oxide can increase the mobility of MoS$_2$ through the purported reduction of CI scattering[17,27,28]. However, another significant, yet often overlooked, factor in the mobility improvement is the high carrier density that can be accessed experimentally in such dual-gated device configurations. In addition, the remote interaction between the electrons and the substrate surface optical (SO) phonons plays a crucial role in limiting charge transport in atomically thin crystals adjacent to high-κ dielectrics[24,29]. Therefore, determining quantitatively the contribution from the various scattering processes to the resistivity poses an enormous challenge that warrants a synergy between an experimental study of the electron mobility with precisely controlled device parameters (including carrier density $n$, dielectric constant $\varepsilon$, density of CI $n_{CI}$, temperature $T$) and numerical modeling to interpret the mobility dependence on these parameters in terms of the underlying scattering mechanisms.

In this Communication, we perform a combined experimental and theoretical study of the electron transport in high-quality, thiol-treated monolayer MoS$_2$ supported on different substrates (SiO$_2$, HfO$_2$ or Al$_2$O$_3$). By suppressing the effects of CI scattering through dielectric and carrier screening, we are able to fabricate monolayer MoS$_2$ transistors with a room-temperature mobility of ~150 cm$^2$/Vs, which is among the highest room-temperature mobility for monolayer MoS$_2$ devices[17,22,23]. Given the excellent sample quality and simple device structure, we can extract quantitatively the contributions from CI, intrinsic and remote phonons. Our analysis confirms that the mobility in these devices is limited by intrinsic and remote phonons rather than CI at room temperature.

Before we report our mobility results, we first give an overview of the physics underlying the scattering of electrons by sources in the substrate. In supported MoS$_2$, the electrons experience random static and time-dependent electric fields at the semiconductor-dielectric interface. These fields are physically created by the CI as well as by the dipoles of the oscillating metal-oxide bonds originating from the polar optical phonons in the dielectric. Therefore, in addition to scattering from the intrinsic acoustic and optical phonons, the interaction between the electrons and these fields introduces another two processes that limit electron mobility: (1) elastic scattering with the CI which are presumably located at the surface of the substrate, and (2) the remote interaction with the SO phonons.

The interaction between the electrons and the CI in the MoS$_2$ can be described by the term $H_{CI} = \sum_{k,q} \rho_{CI}(q) \phi_q^{scr} c_{k+q}^\dagger c_k$ where $\rho_{CI}(q)$ and $\phi_q^{scr}$ are the Fourier transforms of the CI distribution and the screened potential, respectively, and $c_k^\dagger$ ($c_k$) is the electron creation (annihilation) operator. We drop the valley and spin indices to simplify the discussion. The screened potential can be expressed as $\phi_q^{scr} = \frac{\phi_q}{\varepsilon_{2D}(q,T)}$, where $\phi_q = \frac{e^2}{(\varepsilon_{box} + \varepsilon_0)q}$ is the bare potential and $e$ is the electron charge; $\varepsilon_{box}$ and $\varepsilon_0$ are in turn the static permittivity of the substrate and vacuum. The screening of the bare CI by the substrate and the free electrons is described by the generalized screening function[30,31] $\varepsilon_{2D}(q,T) = 1 + \frac{2\varepsilon_{el}(q)}{\varepsilon_{box} + \varepsilon_0}$, where $\varepsilon_{el}(q)$ corresponds to the electronic part of the dielectric function and depends on the carrier density $n$. As $n$ increases, screening becomes stronger, reducing the scattering potential and increasing the CI-limited mobility.

To illustrate the effect of screening, we plot in Fig. 1a the real space potential profile for a point CI in MoS$_2$ under four different scenarios. We observe a reduction in the size of the potential profile when there is polarization charge screening. The effective size of the CI also depends on the substrate dielectric constant which directly reduces the bare potential as well as indirectly weakens polarization charge screening. Therefore, a combination of high κ and high carrier density is most effective in reducing the effective size of the CI, as shown in the lower right panel of Fig. 1a. However, on the HfO$_2$ substrate, despite the large increase in carrier density, the reduction in the profile is not substantial. This is because a high permittivity weakens all Coulomb interactions, including screening.

The variation in κ and $n$ also affects the CI-limited mobility $\mu_{CI}$. Figure 1b shows the simulated $\mu_{CI}$ for the SiO$_2$ substrate over $T$ = 10-400 K and $n$ = 0.1-10×10$^{12}$ cm$^{-2}$ with $n_{CI}$ = 1×10$^{12}$ cm$^{-2}$. The $\mu_{CI}$ increases monotonically with $n$ because screening of CI becomes more effective at higher carrier density. Given that CI scattering is the dominant scattering mechanism in most MoS$_2$ samples, this indicates that the experimental mobility should be measured over a wide range of $n$ before a meaningful comparison can be made with numerical models and that a higher carrier density is required to reach greater mobility. A less intuitive result is the decrease in $\mu_{CI}$ as a function of $T$, which is due to the weakening of the charge polarizability at higher temperatures and is often interpreted mistakenly as a signature of phonon-limited charge transport. This $\mu_{CI}$ decrease has also been shown to be proportionally smaller for a high-κ substrate[30]. Fig. 1c shows the simulated room temperature $\mu_{CI}$ at experimentally accessible $n$ = 3×10$^{12}$, 7×10$^{12}$, 1.1×10$^{13}$ and 1.5×10$^{13}$ cm$^{-2}$ as a function of κ. Generally, $\mu_{CI}$ increases with κ because the higher permittivity reduces the effective charge on the CI and hence the scattering rate. For κ = 40, $\mu_{CI}$ can be as high as ~1100 cm$^2$/Vs at room temperature. However, the extent $\mu_{CI}$ increases with κ is proportionally smaller at high $n$ because the greater screening by the polarization charge diminishes the screening by the substrate (Fig. 1d).

The other major external process that affects the electron mobility is its remote interaction with the substrate SO phonons, described by the term $H_{SO} = \sum_{k,q} M_q c^{\dagger}_{k+q} c_k (a_q + a_{-q})$ where $M_q$ is the coupling coefficient and $a^{\dagger}_q$ ($a_q$) is the phonon creation (annihilation) operator. For a highly polar oxide like HfO$_2$, the coupling coefficient is especially strong because of the large difference between the optical and static dielectric responses. The strength of the bare coupling coefficient can be quantified by the dimensionless coupling constant[32]

$$\alpha_{SO} = \frac{e^2}{8\pi\hbar}\left(\frac{m^*}{2\hbar\omega_{SO}}\right)^{1/2}\left(\frac{1}{\varepsilon^{\infty}_{SO}} - \frac{1}{\varepsilon^{0}_{SO}}\right)$$

, where $\varepsilon^{\infty}_{SO}$ and $\varepsilon^{0}_{SO}$ are the optical and static dielectric response of the interface, respectively. $\alpha_{SO}$ is analogous to the dimensionless Froehlich coupling constant usually defined for polarons in bulk polar insulators and is given in Table 1. Generally, $\alpha_{SO}$ is larger for strongly polar (high-κ) dielectrics such as HfO$_2$ and shares the same physical origin as the high permittivity of the dielectric. In a highly polar insulator, the bonds can be polarized more easily in response to an external electric field and screen a CI more effectively. However, the large polarization from the oscillating metal-oxide bond in the dielectric also couples the associated lattice vibration (phonon) more strongly to free carriers at the surface[33]. As expected, HfO$_2$ (SiO$_2$) has the largest (smallest) average $\alpha_{SO}$ of the three oxides studied in this work and the strongest (weakest) remote phonon scattering as well as the lowest (highest) SO phonon-limited mobility $\mu_{SO}$, shown in Fig. S2. In our semi-classical charge transport model, CI- and SO-phonon scattering are included along with other intrinsic phonon scattering processes. The detailed calculation of mobility is described in Supplementary Information.

The experimental data are obtained from field effect transistors fabricated on high-quality $MoS_2$ samples which we recently developed using thiol chemistry to improve the sample and interface quality[21]. The room-temperature mobility for backgated devices on $SiO_2$, which was limited by CI scattering, could be as high as 80 cm$^2$/Vs. Here we use the same thiol treatment on monolayer $MoS_2$ samples mechanically exfoliated on 10nm high-κ oxide/285nm $SiO_2$/Si substrates (Fig. 2a inset, detailed device fabrication process is described in Supplementary Information). Compared to bare $SiO_2$/Si substrates, the addition of a thin layer of high-κ oxide only changes the gate capacitance $C_g$ by less than 1%, but yields a 50% increase in carrier density by sustaining a much higher (~150V) backgate voltage $V_g$. The dielectric constant of $HfO_2$ and $Al_2O_3$ used in this work are ~16.5 and 10 respectively obtained from standard capacitance measurements (see Supplementary Information). Compared to the topgate devices, our devices are free from the impurities and contaminations introduced in the topgate fabrication[34,35], and much easier to model quantitatively.

Fig. 2a shows variable-temperature measurements of four-probe conductivity $\sigma$ as a function of $V_g$ for a representative device on $HfO_2$ (H1). The curves all intersect near $V_g \approx 70V$ (corresponding to $n=C_gV_g \approx 5.0 \times 10^{12}$cm$^{-2}$), a signature of metal-insulator transition (MIT) due to charge traps at the interface[21]. Since the effect of traps is dominant at low carrier density (comparable or lower than the critical density of MIT) and low temperature[21], we focus our discussion on the opposite limit ($n>7\times 10^{12}$cm$^{-2}$) in order to unambiguously model CI and phonon scattering without being complicated by traps. Fig. 2b shows the four-probe field-effect mobility $\mu(n,T) = \frac{d\sigma(n,T)}{C_g dV_g}$ as a function of temperature for three representative devices on $HfO_2$, $Al_2O_3$ and $SiO_2$ respectively (H1, A2 and S1). For fair comparison, $\mu$ is extracted at $n=7.1\times10^{12}cm^{-2}$ for all three devices. We clearly observe that $\mu$ increases with the dielectric constant of the oxides. The solid lines in Fig. 2b are best theoretical fits at high temperatures taking into account intrinsic and SO phonons and CI. The extracted $n_{CI}$ is similar for all three oxide substrates (Table 1), which is not surprising considering that the oxides have similar roughness (Fig. S3) and are subjected to the same thiol treatment prior to exfoliation of $MoS_2$. However, process variations could cause a fluctuation in $n_{CI}$ of up to ~50%. Since $HfO_2$ ($SiO_2$) has the highest (lowest) average $\alpha_{SO}$ among the three oxides (Table 1), the observed trend in mobility suggests that CI scattering in $HfO_2$ ($SiO_2$) is the weakest (strongest). Indeed, our modeling shows that $\mu_{CI}$ at room temperature increases by 270% when switching $SiO_2$ to $HfO_2$ with similar $n_{CI}$ ~$0.9\times10^{12}cm^{-2}$ (Table 1). This is the first clear demonstration of the dielectric screening effect under well-controlled conditions, unlike in dual-gated devices where CI density is likely increased by the fabrication process and no rigorous comparison can be made with single-gated devices.

Fig. 2c and Fig. S5 depict the temperature dependence of mobility at $n=10.5\times10^{12}cm^{-2}$ for several devices on $HfO_2$ and $Al_2O_3$ respectively. Excellent agreement between experiment (symbols) and modeling (solid lines) is achieved through the whole temperature range of 20-300K, reassuring the accuracy of our model. The device H1 (A2), which is the best device on $HfO_2$ ($Al_2O_3$), shows a room-temperature mobility of 148cm$^2$/Vs (113cm$^2$/Vs), an 85% (41%) improvement over similar samples on $SiO_2$[21]. The mobility at 20K of 847cm$^2$/Vs (591cm$^2$/Vs) shows an even more dramatic improvement. These improvements result primarily from dielectric and carrier screening effects. Using the fitting parameters in Table 1, the contribution of CI and phonons can be quantitatively calculated (dashed and dotted lines Fig. 2c). We find that the lines cross at $T$=233K ($T$=275K) for device H1 (H2), which means that above that temperature, these devices are no longer limited by CI scattering, but by phonons. For device H1, the room-temperature $\mu_{CI}$ and phonon-limited mobility $[\mu_{ph}^{-1}+\mu_{SO}^{-1}]^{-1}$ are 372cm$^2$/Vs and 215cm$^2$/Vs respectively. This is the first time that phonon-limited transport regime has been explicitly demonstrated for any monolayer TMDs. However, for devices on Al$_2$O$_3$, transport is still limited by CI in the measured temperature range due to reduced dielectric screening (Fig. S5b). Furthermore, we find that the experimentally extracted $\mu_{CI}$ can be well fitted by $T^\gamma$ for T>100K, where γ=0.4 and 0.8 for H1 and A2, respectively (Fig. S6b). The power-law dependence of $\mu_{CI}$ is in good agreement with theoretical simulations (Fig. S6a), further supporting our analysis. The smaller γ for HfO$_2$ is attributed to more significant effect of dielectric screening induced by higher dielectric constant.

Recently, it was suggested theoretically that high-κ dielectrics are not ideal substrates for ultra-clean MoS$_2$ because of remote phonon scattering[26]. Our calculation concurs with this proposition. Ultimately, the effective dielectric screening in a high-κ dielectric shares the same physical origins as the strong remote phonon coupling ($\alpha_{SO}$): both are due to the large ionic polarizability of the metal-oxide bond. When the density of CI is reduced, low-κ dielectrics with less polar nature (such as BN and SiO$_2$) will be advantageous. Fig. 2d plots the calculated room-temperature mobility as a function of $n_{CI}$ for the three oxides studied in this work, along with our experimental data. Two regimes emerge in the diagram. At high CI density, the mobility is limited by CI scattering, thus HfO$_2$ is the best among the three oxides due to dielectric screening. A $1/n_{CI}$ scaling is observed in the limit of very strong impurity, where the mobility is generally below 100cm$^2$/Vs. Most experimental results to date fall in this regime[16-23]. At low CI density, the mobility is limited by phonons and is nearly independent of $n_{CI}$. In this regime, HfO$_2$ is outperformed by the other two oxides. If $n_{CI}$ can be reduced below ~0.3×10$^{12}$ cm$^{-2}$ (which is roughly the crossover point between the two regimes), the use of low-κ dielectrics such as SiO$_2$ and BN would be favorable. In this case, one would expect a room-temperature mobility of over 200 cm$^2$/Vs for monolayer MoS$_2$.

Let us now discuss another important aspect of screening by charge carriers, which is manifested experimentally in the carrier-density-dependent mobility. In Fig. 3a we plot the mobility of device H1 under $n$=10.5×10$^{12}$, 8.5×10$^{12}$ and 7.1×10$^{12}$ cm$^{-2}$, in line with the modeling results using the parameters in Table 1 (solid lines). The contribution from phonons (blue shaded region) and CI (yellow shaded region) are plotted separately. The monotonic increase of $\mu$ with $n$ is due to the screening effect of both $\mu_{CI}$ and $\mu_{SO}$. After subtracting the contribution of phonons, we find that $\mu_{CI}$ has a linear relationship with $n$ for all the devices (Fig. 3b). This is because at high temperatures, the polarization charge is diluted so that screening is significantly weaker and the scattering cross section of the CI is similar to that of the bare CI, especially when the substrate κ is high. For a 2D electron gas with a parabolic dispersion, it is known that scattering with a bare Coulomb potential leads to a mobility proportional to $n$.[36] In Fig. 3b, the mobility value and the linear coefficient vary among devices due to different $n_{CI}$ and κ. The variations of $n_{CI}$ can be normalized by $\sigma_{CI} \equiv en_{CI}\mu_{CI}$, which has a unit of conductivity (Fig. 3b inset). After the normalization, devices on the same high-κ substrate collapse onto the same curve, well described by our modeling (lines in Fig. 3b inset). The curve for $HfO_2$ is higher than the one for $Al_2O_3$ because the former's larger dielectric constant reduces the effective charge on its CI, resulting in a proportionally smaller scattering cross section. This consistency provides further evidence that we have accurately differentiated the CI contribution to the resistivity from the phonon contribution.

Another reason for the improved mobility is the screening-induced reduction in remote phonon scattering and $\mu_{SO}$ (Fig. 3a, dotted lines). Recall that the SO phonons are associated with the oscillating polarized metal-oxide bonds in the dielectric. However, the large polarization of the bond also couples the SO phonon to the surface[21] charge from the $MoS_2$, giving rise to screening of the electron-phonon coupling and a smaller contribution to the resistivity.

In conclusion, by systematic engineering of the material quality, dielectric environment and carrier density, we are able to achieve room-temperature phonon-limited transport in monolayer $MoS_2$ for the first time by screening of CI scattering. Through rigorous theoretical modeling, we identify CI and remote phonons as the key limitations in current $MoS_2$ devices. Our model indicates that there is limited room for further mobility improvement on $HfO_2$ ($\mu$<215 cm$^2$/Vs) if we are constrained by the possibility of dielectric breakdown to a maximum carrier density of $1\times10^{13}$ cm$^{-2}$. The present methodology of combining high-k dielectric screening and interface functionalization is a generic route to increase carrier mobility in other TMDs such as $WS_2$ (Ref. 37). Future improvement of the mobility in TMDs requires continued interface engineering that combines a low CI density and weak remote phonon scattering simultaneously.

**Supporting Information**

Supporting Information is available from the Wiley Online Library or from the author.

Acknowledgements
Z. H. Yu, Z. Y. Ong, Y. M. Pan contributed equally to this work. This work was supported in part by National Key Basic Research Program of China 2013CBA01604, 2015CB921600; National Natural Science Foundation of China 61325020, 61261160499, 11274154, 61521001; MICM Laboratory Foundation 9140C140105140C14070; a project funded by the Priority Academic Program Development of Jiangsu Higher Education Institutions; "Jiangsu Shuangchuang" program and 'Jiangsu Shuangchuang Team' Program.

Received: ((will be filled in by the editorial staff))
Revised: ((will be filled in by the editorial staff))
Published online: ((will be filled in by the editorial staff))

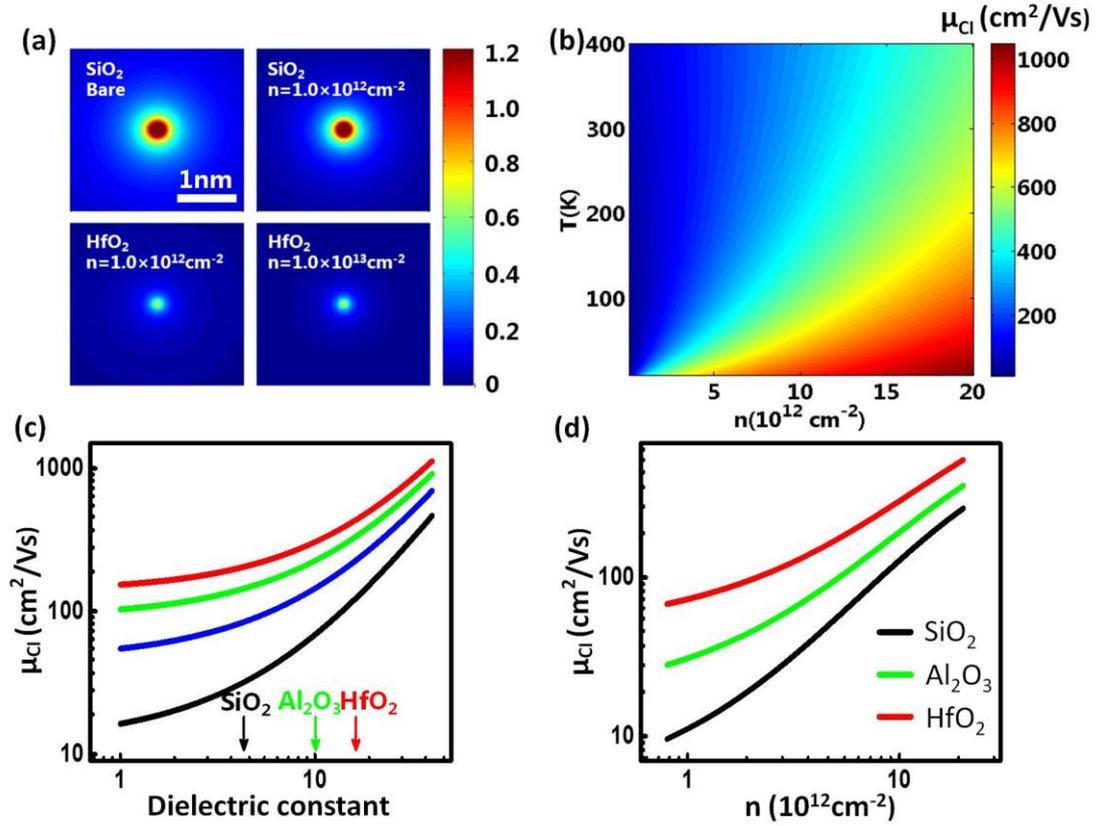

**Figure 1. Modeling of charged impurity scattering in monolayer MoS$_2$.** (a) The real space distribution of screened Coulomb potential for a point charge in MoS$_2$ on SiO$_2$ ($\kappa$ = 3.9, upper panels) and HfO$_2$ ($\kappa$ = 16.5, lower panels) at different carrier densities. (b) $\mu_{CI}$ as a function of carrier density and temperature on HfO$_2$ substrate ($n_{CI}$=1.0×10$^{12}$ cm$^{-2}$). (c) $\mu_{CI}$ as a function of dielectric constant at different carrier density (*T*=300K). From top to bottom, *n*=1.5×10$^{13}$ cm$^{-2}$ (red), 1.1×10$^{13}$ cm$^{-2}$ (green), 7.0×10$^{12}$ cm$^{-2}$ (blue) and 3.0×10$^{12}$ cm$^{-2}$ (black) respectively. (d) $\mu_{CI}$ as a function of carrier density on SiO$_2$ (black), Al$_2$O$_3$ (green) and HfO$_2$ (red) substrates (*T*=300K, $n_{CI}$=1.0×10$^{12}$cm$^{-2}$).

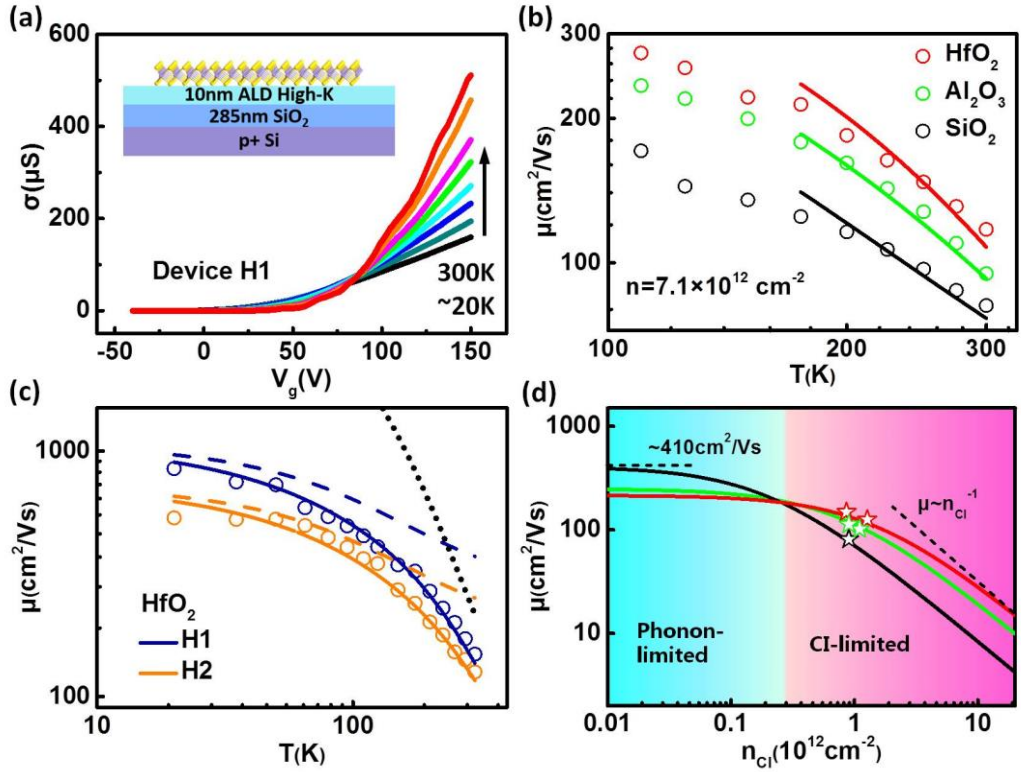

**Figure 2. Effect of dielectric screening on the charge transport of monolayer MoS$_2$.** (a) Four-probe conductivity as a function of $V_g$ for a representative device on HfO$_2$ substrate (device H1). Inset shows a cartoon illustration of the device structure. (b) Field-effect mobility as a function of temperature for three devices on SiO$_2$ (black), Al$_2$O$_3$ (green) and HfO$_2$ (red) respectively, under $n=7.1\times10^{12}$ cm$^{-2}$ (symbols). Solid lines are the modeling results at high temperature. (c) Field-effect mobility as a function of temperature for two devices on HfO$_2$ substrate under $n=10.5\times10^{12}$ cm$^{-2}$ (symbols), together with the best theoretical fittings (solid lines, see Table 1 for fitting parameters), the calculated CI-limited mobility (dashed lines), and the calculated phonon-limited mobility (black dotted line). (d) Predicted field-effect mobility as a function of $n_{CI}$ for devices on SiO$_2$ (black line), Al$_2$O$_3$ (green line) and HfO$_2$ (red line)

substrate. The symbols are experimentally extrapolated data from five different devices (the parameters are listed in Table 1).

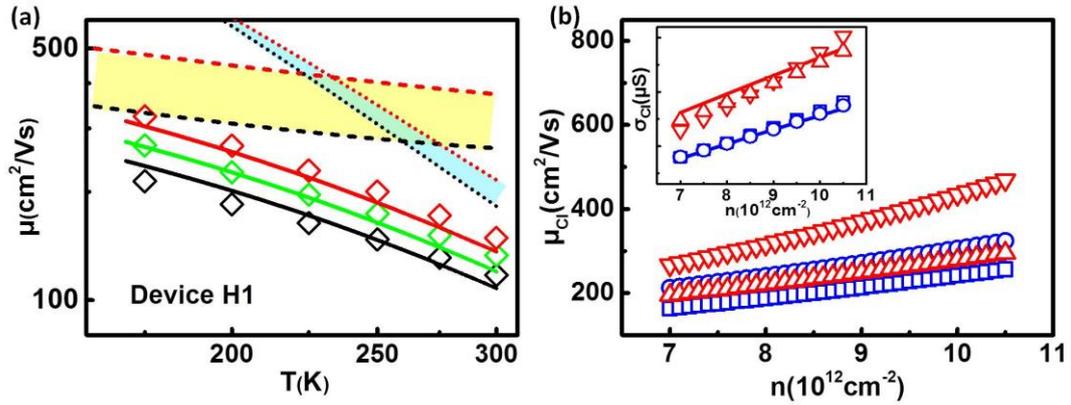

**Figure 3. Effect of carrier screening on the charge transport of monolayer MoS$_2$.** (a) Field-effect mobility as a function of temperature under different carrier densities for device H1 on HfO$_2$ (symbols), together with the theoretical fittings (solid lines, see Table 1 for fitting parameters). From top to bottom, $n=10.5\times10^{13}$ cm$^{-2}$ (red), $8.5\times10^{12}$ cm$^{-2}$ (green) and $7.1\times10^{12}$ cm$^{-2}$ (black) respectively. The calculated CI-limited mobility and phonon-limited mobility in the same carrier density regime are denoted by the yellow and blue shaded area respectively. (b) CI-limited mobility as a function of carrier density for the four devices on Al$_2$O$_3$ and HfO$_2$ at $T$=200K. Red up triangles: device H1; red down triangles: device H2; blue squares: device A1; blue circles: device A2. Inset is the normalized conductivity $\sigma_{CI}$ as a function of carrier density for the four devices. Solid lines are modeling results using the parameters in Table 1.

| Oxide | $\varepsilon$ | $\varepsilon^\infty$ | $\alpha_{SO1}$ | $\alpha_{SO2}$ | $\omega_{TO1}$ (meV) | $\omega_{TO2}$ (meV) | $\omega_{LO1}$ (meV) | $\omega_{LO2}$ (meV) | $\omega_{SO1}$ (meV) | $\omega_{SO2}$ (meV) | Device # | $n_{Cl}$ (cm$^{-2}$) | $\mu_{Cl}$ (300K) (cm$^2$/Vs) | $\mu$ (300K) (cm$^2$/Vs) |
|---|---|---|---|---|---|---|---|---|---|---|---|---|---|---|
| HfO$_2$ | 16.5 | 4.23 | 1.27 | - | 40.0 | - | 79.0 | - | 73.2 | - | H1 | 0.86×10$^{12}$ | 372 | 148 |
|  |  |  |  |  |  |  |  |  |  |  | H2 | 1.27×10$^{12}$ | 252 | 125 |
| Al$_2$O$_3$ | 10 | 2.56 | 0.18 | 1.32 | 48.2 | 71.4 | 56.5 | 120.4 | 56.0 | 108.0 | A1 | 1.1×10$^{12}$ | 164 | 101 |
|  |  |  |  |  |  |  |  |  |  |  | A2 | 0.85×10$^{12}$ | 246 | 113 |
| SiO$_2$ | 3.9 | 2.5 | 0.41 | 0.28 | 55.6 | 138.1 | 62.6 | 153.3 | 61.0 | 149.0 | S1 | 0.9×10$^{12}$ | 101 | 81 |

**Table 1** The mobility values and fitting parameters of the devices presented in this work are taken and derived from Refs. 27, 29 and 31.

**Supplementary Information**

1. Details of theoretical calculations

2. Growth and characterizations of high-κ oxides

3. Sample preparation, device fabrication and electrical measurements

4. Additional electrical data of devices on Al$_2$O$_3$ and HfO$_2$ substrate

1. **Details of theoretical calculations**

To analyze the electron mobility in monolayer MoS$_2$, we use a semiclassical model based on the relaxation time approximation. The electrons are assumed to move diffusively in the MoS$_2$ sheet, with an effective mass[1] of m* = 0.48m$_0$ where m$_0$ is the free electron mass. However, their mobility is limited by scattering processes intrinsic and extrinsic to MoS$_2$. The expression for the electron mobility is given by[2, 3]

$$\mu = \frac{2e}{\pi n \hbar^2 k_B T} \int_0^\infty f(E)[1-f(E)]\Gamma(E)^{-1} E dE \tag{S1}$$

where $e$, $n$, $\hbar$, $k_B$ and $T$ are the electron charge quantum, the electron density, the Planck constant divided by 2π, the Boltzmann constant and the temperature, respectively. The functions $f(E)$ and $\Gamma(E)$ in turn represent the Fermi-Dirac distribution and the momentum relaxation rate. The total electron mobility is computed using Matthiesen's rule[2]: $\mu = (\mu_{CI}^{-1} + \mu_{ph}^{-1} + \mu_{SO}^{-1})^{-1}$ where $\mu_{CI}$, $\mu_{ph}$ and $\mu_{SO}$ are the individual mobilities limited by scattering with CI, intrinsic phonons or SO phonons, respectively, and calculated using Eq. (S1) with the scattering rates $\Gamma_{CI}$ (Coulomb impurities), $\Gamma_{ph}$ (intrinsic phonons) or $\Gamma_{SO}$ (surface optical phonons). The details of how the scattering rates are given in the following discussions.

a. **Coulomb impurities**

Coulomb impurity scattering is due to the charge centers near the oxide surface and also to sulfur vacancies within the MoS$_2$, and is known to be a major source of resistance to electron conduction in 2D crystals. It was shown by Ong and Fischetti[3] that CI-limited charge transport can have an electron mobility temperature dependence that is similar to that in phonon-limited charge transport.

The screened potential can be expressed as $\phi_q^{scr} = \frac{\phi_q}{\varepsilon_{2D}(q,T)}$, where $\phi_q = \frac{e^2}{(\varepsilon_{box}^0 + \varepsilon_0)q}$ is the bare potential; $\varepsilon_{box}^0$ and $\varepsilon_0$ are in turn the static permittivity of the substrate and vacuum. The screening of the bare CI by the substrate and the free electrons is described by the generalized screening function $\varepsilon_{2D}(q,T) = 1 + \frac{2\varepsilon_{el}(q)}{\varepsilon_{box}^0 + \varepsilon_0}$, where $\varepsilon_{el}(q) = -\frac{e^2}{2q}\Pi(q,T,E_F)$ corresponds to the electronic part of the dielectric function. $\Pi(q,T,E_F)$ is the temperature- and carrier density-dependent static polarizability, and represents the polarization charge screening of the CI. The exact form of $\Pi(q,T,E_F)$ is given in Refs. 3-5.

The CI scattering rate is[3]

$$\Gamma_{CI}(E_k) = \frac{n_{CI}}{2\pi\hbar} \int dk \, |\phi_{|k-k'|}^{scr}|^2 \, (1-\cos\theta_{kk'})\delta(E_k - E_{k'}), \qquad (S2)$$

b. where $\theta_{kk'}$ is the scattering angle between the $k$ and $k'$ states and $E_k$ is the energy. Thus, the CI-limited mobility $\mu_{CI}$ can be calculated using Eqs. (S1) and (S2).**Intrinsic phonon scattering**

In our model, the intrinsic phonon scattering rate ($\Gamma_{ph}$) is due to electrons scattering with the longitudinal (LA) and transverse acoustic (TA), the intravalley polar longitudinal optical (Froehlich), the intervalley polar longitudinal optical (LO) and the intravalley homopolar optical (HP) phonons. Thus, we have $\Gamma_{ph} = \Gamma_{LA} + \Gamma_{TA} + \Gamma_{LO} + \Gamma_{HP} + \Gamma_{Fr}$ where

$\Gamma_{LA}$, $\Gamma_{TA}$, $\Gamma_{LO}$, $\Gamma_{HP}$ and $\Gamma_{Fr}$ are the scattering rates associated with LA, TA, LO, HP and Fr. phonons.

The LA and TA phonon scattering rates are given by $\Gamma_{ac} = \frac{m^* \Xi_{ac}^2 k_B T}{\hbar^3 \rho c_{ac}^2}$, where $\Xi_{ac}$ is the acoustic phonon (LA or TA) deformation potential and $c_{ac}$ is the acoustic phonon speed.

The intervalley polar longitudinal optical (LO) and intravalley homopolar optical (HP) phonon scattering rates are given by $\Gamma_{op}(E) = \frac{m^* D_{op}^2}{2\hbar^2 \rho \omega_{op}}[N_{op} + (1 + N_{op})\Theta(E - \hbar\omega_{op})]$, where $D_{op}$ is the optical deformation potential of the optical phonon (LO or HP) and $N_{op} = [\exp(\hbar\omega_{op}/k_V T) - 1]^{-1}$ is its Bose-Einstein distribution with $\omega_{op}$ its phonon energy. $\Theta(\ldots)$ is the usual Heaviside function.

In the long wavelength limit, the intravalley polar longitudinal optical (Froehlich) phonons can couple with the electron gas and undergo screening. Thus, the Froehlich optical phonon emission (+) or absorption (-) rate is[4]:

$$\Gamma_{Fr}^{\pm}(E) = \frac{e^2 \omega_{Fr} m^*}{8\pi\hbar^2}\left(\frac{1}{2} \pm \frac{1}{2} + N_{Fr}\right)\int_{-\pi}^{\pi} d\theta \frac{1 - \frac{k'}{k}\cos\theta}{q}\left(\frac{1}{\varepsilon_{ion}^{\infty} + \varepsilon_{el}(q)} - \frac{1}{\varepsilon_{ion}^{0} + \varepsilon_{el}(q)}\right)\text{erfc}(\frac{q\sigma}{2})^2$$

where $k$ is the initial state, $k'$ is the final state given by $k' = \sqrt{k^2 \mp 2m^*\omega_{Fr}/\hbar}$, and $q = \sqrt{k^2 + k'^2 - 2kk'\cos\theta}$. Also, $\varepsilon_{ion}^{\infty}$ ($\varepsilon_{ion}^{0}$) is the ionic part of the optical (static) permittivity of MoS$_2$, erfc is the complementary error function and $\sigma$ is the sheet thickness.. We assume the phonon dispersion for the LO phonons to be flat so that $\omega_{Fr} = \omega_{LO}$. Therefore, the Froehlich phonon scattering rate is:

$\Gamma_{Fr}(E) = \Gamma_{Fr}^{-}(E) + \Theta(E - \hbar\omega_{Fr})\Gamma_{Fr}^{+}(E)$.

Fig. S1a shows the intrinsic phonon-limited mobility $\mu_{ph}$ as a function of temperature at n=10$^{13}$ cm$^{-2}$, calculated using the parameters in Table S1. $\mu_{ph}$ decreases with temperature and reaches ~420 cm$^2$/Vs at 300 K which is the maximum room-temperature electron mobility if there is no scattering with CI and SO phonons and in good agreement with published results[1, 4]. Fig. S1b shows $\mu_{ph}$ as a function of carrier density at 300 K. As we increase n, $\mu_{ph}$ grows initially because of the screening of the Froehlich coupling reduces its contribution to the intrinsic phonon scattering rate. However, at higher n, the scattering rate increases because the Fermi level is at the higher energy states where there is greater optical phonon (LO, HP and Fr.) emission and the corresponding scattering rate increases.

| Parameter | Numerical value |
|---|---|
| $m^*$ | $0.48\, m_0$ |
| $\Xi_{LA}$ ($\Xi_{TA}$) | 2.8 eV (1.6 eV) |
| $c_{LA}$ ($c_{TA}$) | 6700 m/s (4200 m/s) |
| $\rho$ | $3.1 \times 10^{-7}$ g/cm$^{-2}$ |
| $\sigma$ | $4.41 \times 10^{-10}$ m |
| $D_{LO}$ ($D_{HP}$) | $2.6 \times 10^8$ eV/cm ($4.1 \times 10^8$ eV/cm) |
| $\omega_{LO}$ ($\omega_{HP}$) | 48 meV (50 meV) |
| $\varepsilon_{ion}^0$ ($\varepsilon_{ion}^\infty$) | $7.6\, \varepsilon_0$ ($7.0\, \varepsilon_0$) |

Table S1. Parameters used for the intrinsic phonon scattering rates. The values are taken from Ref. 1, 6.

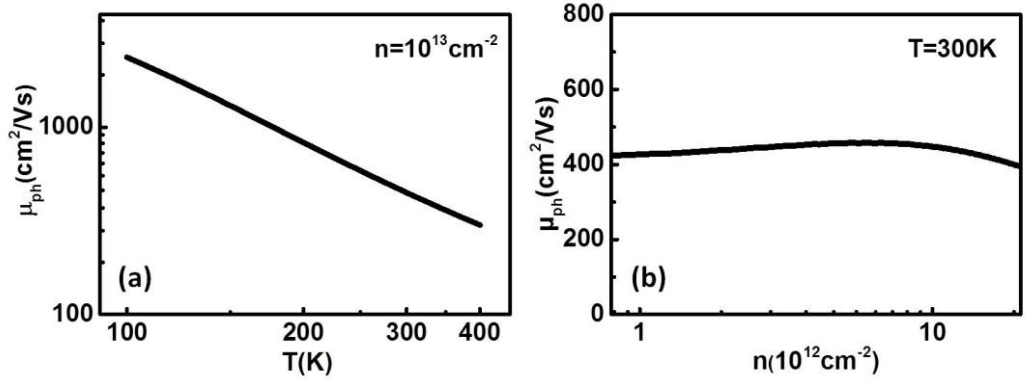

Figure S1. Theory for intrinsic phonon limited mobility. (a) $\mu_{ph}$ as a function of temperature at n=1.0×10$^{13}$cm$^{-2}$. (b) $\mu_{ph}$ as a function of carrier density at T=300K.

### c. Surface optical phonon scattering

Another source of electron scattering is through remote interaction with the polar optical phonons in the dielectric substrate[4,7]. We assume that there are two phonon modes in SiO$_2$ and Al$_2$O$_3$. The SO phonon scattering rate is given by $\Gamma_{SO} = \Gamma_{SO1} + \Gamma_{SO2}$. For HfO$_2$, we assume that there is only one phonon mode though like in Ref.[8].

The dielectric function of the substrate $\varepsilon_{box}(\omega)$ is[7,9]:

$$\varepsilon_{box}(\omega) = \varepsilon_{box}^{\infty} + (\varepsilon_{box}^{i} - \varepsilon_{box}^{\infty})\frac{\omega_{TO2}^2}{\omega_{TO2}^2 - \omega^2} + (\varepsilon_{box}^{0} - \varepsilon_{box}^{i})\frac{\omega_{TO1}^2}{\omega_{TO1}^2 - \omega^2}$$

where $\varepsilon_{box}^{0}$, $\varepsilon_{box}^{i}$ and $\varepsilon_{box}^{\infty}$ are the static, intermediate and optical dielectric of the substrate, respectively, and $\omega_{TO1}$ and $\omega_{TO2}$ are the transverse optical phonon angular frequencies such that $\omega_{TO1} < \omega_{TO2}$. We can rewrite $\varepsilon_{box}(\omega)$ in the generalized Lyddane-Sachs-Teller form

$$\varepsilon_{box}(\omega) = \varepsilon_{box}^{\infty}\left(\frac{\omega_{LO1}^2 - \omega^2}{\omega_{TO1}^2 - \omega^2}\right)\left(\frac{\omega_{LO2}^2 - \omega^2}{\omega_{TO2}^2 - \omega^2}\right)$$

. In the case of HfO$_2$, we consider only one TO mode, so that its dielectric function is

$$\varepsilon_{box}(\omega) = \varepsilon_{box}^{\infty} + (\varepsilon_{box}^{0} - \varepsilon_{box}^{\infty})\frac{\omega_{TO}^2}{\omega_{TO}^2 - \omega^2} = \varepsilon_{box}^{\infty}\left(\frac{\omega_{LO}^2 - \omega^2}{\omega_{TO}^2 - \omega^2}\right) \quad \text{while} \quad \varepsilon_{tot,SO}^{\infty} = \frac{1}{2}\left(\varepsilon_{box}^{\infty} + \varepsilon_{0}\right)$$

and $\varepsilon_{tot,SO}^{0} = \frac{1}{2}\left[\varepsilon_{box}^{\infty}\left(\frac{\omega_{LO}^2}{\omega_{TO}^2}\right) + \varepsilon_{0}\right]$. The surface optical phonon frequencies ($\omega_{SO1}$ and $\omega_{SO2}$) can be determined from the roots of the equation $\varepsilon_{box}(\omega) + \varepsilon_{0} = 0$.

The remote interaction with the substrate SO phonons is described by the term $H_{SO} = \sum_{k,q} M_q c_{k+q}^{\dagger} c_k (a_q + a_{-q}^{\dagger})$ where $M_q$ is the coupling coefficient and $a_q^{\dagger}$ ($a_q$) is the phonon creation (annihilation) operator. The coupling coefficient is given by

$$M_q = \left[\frac{e^2 \hbar \omega_{SO}}{\Omega q}\left(\frac{1}{\varepsilon_{SO}^{\infty} + \varepsilon_{el}(q)} - \frac{1}{\varepsilon_{SO}^{0} + \varepsilon_{el}(q)}\right)\right]^{1/2}$$

where $\omega_{SO}$ is the SO phonon energy (Table 1). $\varepsilon_{SO}^{\infty}$ and $\varepsilon_{SO}^{0}$ are the optical and static dielectric response of the interface, respectively. The term $\varepsilon_{el}(q)$ represents the screening effect of the electron gas on the surface electric field of the substrate SO phonons and like in CI scattering, depends on the temperature as well as the carrier density.

The remote optical phonon emission (+) and absorption (-) rates (SO1) can be written as

$$\Gamma_{SO1}^{\pm}(E) = \frac{e^2 \omega_{SO1} m^*}{8\pi \hbar^2}\left(\frac{1}{2} \pm \frac{1}{2} + N_{SO1}\right)\int_{-\pi}^{\pi} d\theta \frac{1 - \frac{k'}{k}\cos\theta}{q}\left(\frac{1}{\varepsilon_{tot,SO1}^{\infty} + \varepsilon_{el}(q)} - \frac{1}{\varepsilon_{tot,SO1}^{0} + \varepsilon_{el}(q)}\right)$$

where $\varepsilon_{tot,SO1}^{\infty} = \frac{1}{2}\left[\varepsilon_{box}^{\infty}\left(\frac{\omega_{LO2}^2 - \omega_{SO1}^2}{\omega_{TO2}^2 - \omega_{SO1}^2}\right) + \varepsilon_{0}\right]$ and

$\varepsilon_{tot,SO1}^{0} = \frac{1}{2}\left[\varepsilon_{box}^{\infty}\left(\frac{\omega_{LO1}^2}{\omega_{TO1}^2}\right)\left(\frac{\omega_{LO2}^2 - \omega_{SO1}^2}{\omega_{TO2}^2 - \omega_{SO1}^2}\right) + \varepsilon_{0}\right]$. Thus, the SO1 scattering rate is:

$\Gamma_{SO1}(E) = \Gamma_{SO1}^{-}(E) + \Theta(E - \hbar\omega_{SO1})\Gamma_{SO1}^{+}(E)$. The SO2 scattering rate is similarly defined.

Figure S2 shows the SO phonon-limited mobility $\mu_{SO}$ calculated using Eq. (S1)

as a function of temperature at different values of n for $SiO_2$, $Al_2O_3$ and $HfO_2$. We observe that $\mu_{SO}$ increases with *n* because the remote interaction with the SO phonons undergoes greater screening at higher *n* like in CI scattering. Also, its decreases with temperature at a greater rate than $\mu_{ph}$. Thus, at room temperature, it is one of the most dominant scattering processes especially for $HfO_2$.

|  | $SiO_2$ | $Al_2O_3$ | $HfO_2$ | $Al_2O_3$ (ref.) | $HfO_2$ (ref.) |
|---|---|---|---|---|---|
| $\omega_{TO1}$ (meV) | 55.60 | 48.18 | 40.0 | 48.18 | 40.0 |
| $\omega_{TO2}$ (meV) | 138.10 | 71.41 | - | 71.41 | - |
| $\omega_{LO1}$ (meV) | 62.57 | 56.47 | 79.00 | 56.47 | 79.0 |
| $\omega_{LO2}$ (meV) | 153.28 | 120.55 | - | 120.55 | - |
| $\omega_{SO1}$ (meV) | 60.99 | 56.00 | 73.17 | - | - |
| $\omega_{SO2}$ (meV) | 148.87 | 108.00 | - | - | - |
| $\varepsilon_{box}^{0}$ ($\varepsilon_0$) | 3.90 | 10.00 | 16.50 | 12.53 | 16.00 |
| $\varepsilon_{box}^{i}$ ($\varepsilon_0$) | 3.05 | 5.80 | - | 7.27 | - |
| $\varepsilon_{box}^{\infty}$ ($\varepsilon_0$) | 2.50 | 2.56 | 4.23 | 3.20 | 4.10 |

Table S2. Parameters used for the SO phonon scattering rates. The values for $SiO_2$ and $Al_2O_3$ (ref.) are taken from Ref. 3; the values for $HfO_2$ (ref.) are taken from Ref. 8. $\varepsilon_{box}^{0}$ (or κ) for $Al_2O_3$ and $HfO_2$ are extracted from our capacitance measurements. The other parameters ($\varepsilon_{box}^{i}$ and $\varepsilon_{box}^{\infty}$) for $Al_2O_3$ and $HfO_2$ are obtained by rescaling the values for $Al_2O_3$ (ref.) and $HfO_2$ (ref.).

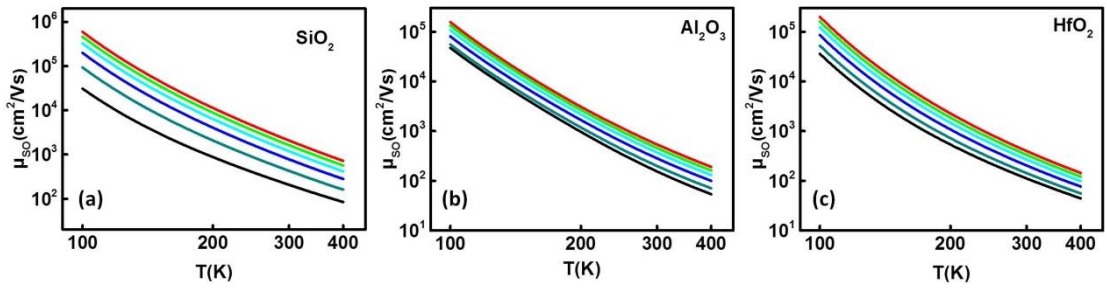

Figure S2. Theory for surface optical phonon limited mobility for different substrates. From top to bottom, n=1.0×10$^{13}$cm$^{-2}$ (red), 8×10$^{12}$cm$^{-2}$ (green), 6×10$^{12}$cm$^{-2}$ (cyan), 4×10$^{12}$cm$^{-2}$ (blue), 2×10$^{12}$cm$^{-2}$ (dark cyan) and 1×10$^{12}$cm$^{-2}$ (black).

## 2. Growth and characterizations of high-κ oxides

We grew 10nm thick high-κ oxide on SiO$_2$/Si substrate by Atomic Layer Deposition (ALD). Before ALD, the substrate was thoroughly cleaned by acetone and isopropanol. For Al$_2$O$_3$, trimethylaluminum (TMA, Micro-nano Tech. Co. Ltd., China) and H$_2$O were used as precursors. For HfO$_2$, tetrakis(dimethylamido)hafnium (TDMAH, Micro-nano Tech. Co. Ltd., China) and H$_2$O were used as precursors. The deposition temperature was maintained at 150 °C for both oxides.

We characterized the oxide films by surface roughness and dielectric constant measurements. The average mean-square roughness (R$_q$) calculated by AFM images is less than 0.2nm (Fig. S3a, b), similar to the SiO$_2$ substrate (Fig. S3c). This confirms the highly uniform and smooth surface of high-κ oxides, which is important for low CI. We measured the dielectric constant of the oxides by standard capacitance measurement with Au/oxide/Au structure using the Agilent E4980A precision LCR meter. The thickness of the oxides is ~100nm as obtained by AFM. We used the capacitance value at 10kHz, due to the negligible change of capacitance in low frequency regime.

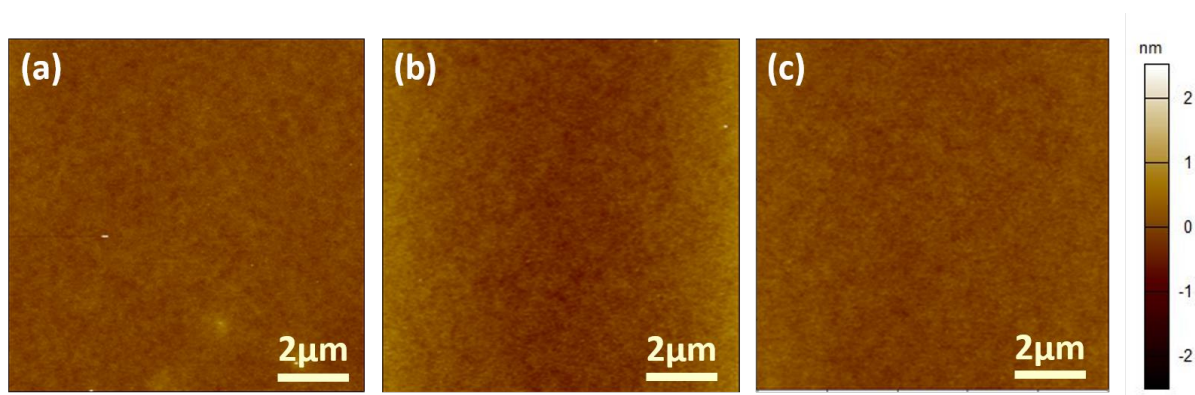

Figure S3. AFM characterization of high-κ oxide films. AFM images of (a) $Al_2O_3$ on $SiO_2$, (b) $HfO_2$ on $SiO_2$ and (c) bare $SiO_2$. The images all have the same Z scale.

3. **Sample preparation, device fabrication and electrical measurements**

We used the double-side MPS treatment reported in Ref. 10 for all samples studied here. Briefly, the 10nm high-κ oxide/285nm $SiO_2$/Si substrate was first subjected to a 30-min UV/ozone treatment to hydroxylate the oxide surface. The substrate was then dipped in a 10% (v/v) MPS/dichloromethane solution for 12 h in a dry glove box to grow MPS self-assembled monolayer. When MPS growth was finished, the substrate was sonicated in dichloromethane followed by thorough rinsing with dichloromethane and IPA, and drying with $N_2$. We then exfoliated monolayer $MoS_2$ from natural flakes (SPI Supplies) on the MPS-treated substrate. After exfoliation, the sample was dipped in a fresh solution of 1/15 (v/v) MPS/dichloromethane for 24 h in a dry glove box to grow MPS on the top side of $MoS_2$, followed by thorough rinsing. Finally, the sample was annealed in a mixture of $H_2$/Ar at 350 °C for 20 min to finish the MPS treatment.

The source, drain and voltage probe electrodes of the field-effect transistors were then patterned by standard electron beam lithography, followed by electron beam evaporation of Ti/Pd (20nm/20nm) and lift-off. In the ebeam lithography step, we use double layer resist stack (MMA/PMMA), to reduce the exposure dose and form the undercut geometry. The devices were then annealed at 350°C in Ar atmosphere for half an hour to improve contacts. Transport measurements were performed by a Keithley 4200 semiconductor parameter analyzer in a close-cycle cryogenic probe station with base pressure ~$10^{-5}$ Torr. 350K in situ vacuum annealing was performed to remove adsorbates and improve device performances before measurement[11].

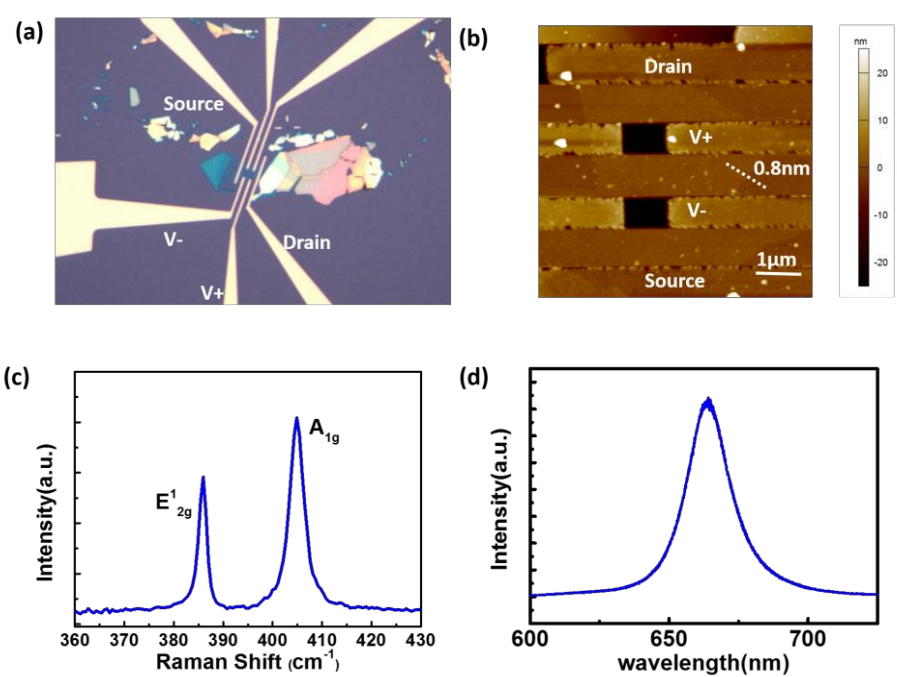

Figure S4. (a) Optical image of monolayer $MoS_2$ devices. (b) AFM image of monolayer $MoS_2$ device with hight of 0.8nm. (c) Raman spectrum of monolayer $MoS_2$ at 514.5 nm excitation.

The spectrum is shown with a distance of 19 cm$^{-1}$ between two vibrating modes (E1$_{2g}$ and A1$_g$)[12]. (d) PL characterizations of monolayer MoS$_2$.

## 4. Additional electrical data of devices on Al$_2$O$_3$ and HfO$_2$ substrate

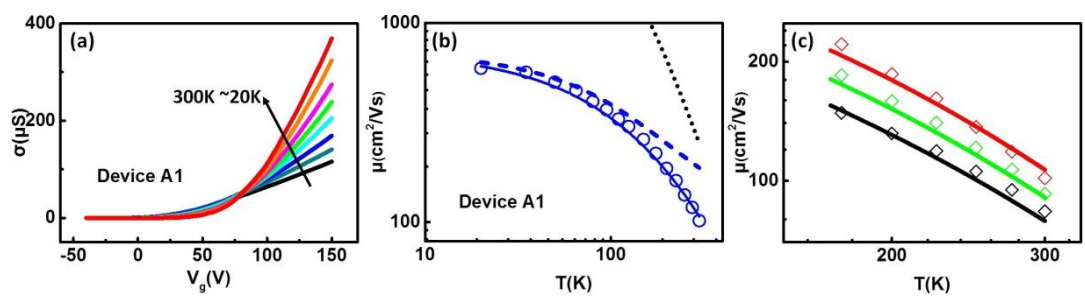

Figure S5. Electrical data and theoretical modeling for device A1 on Al$_2$O$_3$ substrate. (a) Four-probe conductivity as a function of V$_g$ from 300K to 20K. (b) Field-effect mobility as a function of temperature under n=1.0×10$^{13}$cm$^{-2}$ (symbols). The solid line is the best theoretical fitting (see Table 1 for fitting parameters). The blue dash line is calculated CI-limited mobility. The black dot line is the calculated phonon-limited mobility. (c) Field-effect mobility as a function of temperature at different carrier densities. The solid lines are the calculated results using parameters in Table 1. From top to bottom, n=1.0×10$^{13}$cm$^{-2}$ (red), 8.5×10$^{12}$cm$^{-2}$ (green) and 7.0×10$^{12}$cm$^{-2}$ (black).

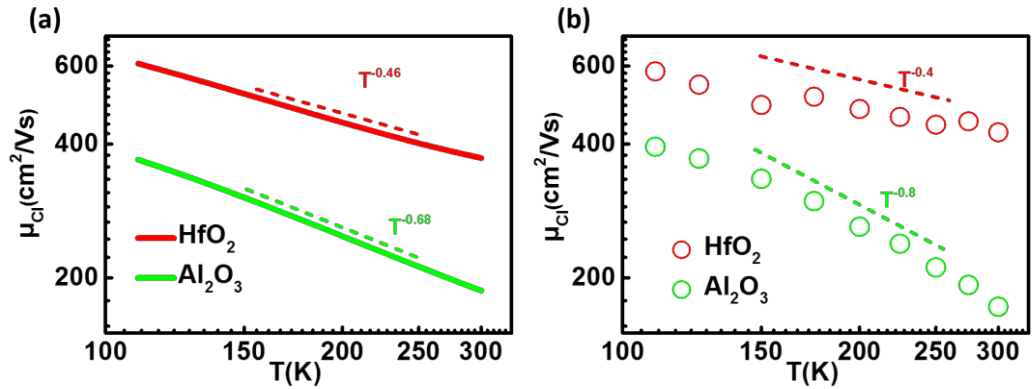

Figure S6. Theoretical (a) and experimental (b) $\mu_{Cl}$ as a function of temperature under n=1.05×10$^{13}$cm$^{-2}$. The dash lines show $T^{\gamma}$ scaling. Theoretically, the $\mu_{Cl}$ can be fitted by $T^{\gamma}$, where γ is an intrinsic parameter that reflects the dielectric screening. γ decreases as dielectric constant increases, γ=0.46 for HfO$_2$ and γ=0.68 for Al$_2$O$_3$. The experimental data follow the same trend as theoretical predictions. The best power law fitting gives γ=0.4 for HfO$_2$ and γ=0.8 for Al$_2$O$_3$, in reasonable agreement with our simulations.